\documentstyle[12pt]{article}
\newcommand{\be}{\begin{equation}}
\newcommand{\ee}{\end{equation}}
\newcommand{\ben}{\begin{eqnarray}}
\newcommand{\een}{\end{eqnarray}}
\begin{document}

\begin{center}
{\bf Travelling Wave Solutions in Nonlinear\\
Diffusive and Dispersive Media\footnote{This work is supported in part
by funds provided by the U. S. Department of Energy (D.O.E.) under
cooperative research agreement DE-FC02-94ER40818, and by Conselho
Nacional de Desenvolvimento Cient\'\i fico e Tecnol\'ogico, CNPq, Brazil.}}
\end{center}

\begin{center}
D. Bazeia\footnote{On leave from Departamento de F\'\i sica, Universidade
Federal da Para\'\i ba, Caixa Postal 5008, 58051-970 Jo\~ao Pessoa,
Para\'\i ba, Brazil}$^{\ast}$ 
and E. P. Raposo$^{\dag}$
\end{center}

\begin{center}
$^{\ast}$Center for Theoretical Physics\\
Laboratory for Nuclear Science and Department of Physics\\
Massachusetts Institute of Technology, Cambridge, Massachusetts 02139-4307\\

$^{\dag}$Lyman Laboratory of Physics\\
Harvard University, 
Cambridge, Massachusetts 02138
\end{center}

\vskip 2cm

\begin{center}
(MIT-CTP-2734, April 1998)
\end{center}

\vskip 2cm

\begin{center}
Abstract
\end{center}

We investigate the presence of soliton solutions in some classes of 
nonlinear partial differential equations, namely 
generalized Korteweg-de Vries-Burgers, Korteveg-de Vries-Huxley, 
and Korteveg-de Vries-Burgers-Huxley equations, which 
combine effects of diffusion, dispersion, and nonlinearity. 
We emphasize the chiral behavior of the travelling solutions, whose
velocities are determined by the parameters that define the equation. 
For some appropriate choices, we show that these equations can be mapped 
onto equations of motion of relativistic $1+1$ dimensional $\phi^{4}$ and
$\phi^{6}$ field theories of real scalar fields. We also study systems of
two coupled nonlinear equations of the types mentioned.

\vskip 1.0cm

\begin{center}
\end{center}

\newpage

\section{Introduction}

\indent
Nonlinear partial differential equations are ubiquitous in 
physics. Their applications range from magnetofluid dynamics to water surface 
gravity waves, electromagnetic radiation reactions, and ion acoustic waves
in plasmas, among several others. In a number of them, it is well known that 
localized travelling solutions, or solitons, appear. Since the 
seminal work of Korteweg and de Vries (KdV) \cite{kdv}, in which a third order 
nonlinear equation was studied to explain the shallow-water 
solitary wave experiments by Russel \cite{rus}, 
high order equations have been used to describe physical phenomena 
occurring in a broad range of fields \cite{whi74,new}, from liquid crystals
to dynamics of growing interfaces and domain walls, also including 
many applications in eletromagnetism, nonlinear optics, acoustics, 
and elasticity. In addition, another interesting possibility arises in some
special cases in which the soliton solutions present chirality, i.e., 
non-arbitrary velocities of propagation 
determined by the parameters of the differential equation, 
and with definite sign. In this context, recent works \cite{ben96,jac96}
have shown that the nonlinear Schr\"odinger equation presents chiral solitons
when nonlinearity enters the game via derivative coupling.
The nonlinear Schr\"odinger equation that appears in these
works is obtained from a very interesting dimensional reduction to
one space dimension of a planar model \cite{jpi90} describing non-relativistic
matter coupled to a Chern-Simons gauge field.
This equation can be cast to the form
\be
iu_t+\lambda\,j\,u+\mu\,u_{xx}=\frac{dV}{d\rho}\,u~,
\ee
where $j=-i\nu(u^{*}u_x -u\,u^{*}_x)$ is the current density, $\lambda,\mu,\nu$
are real parameters and $V=V(u^*u)=V(\rho)$ is the potential, expressed in
terms of the charge density. This equation should be contrasted with
\be
iu_t+\mu\,u_{xx}=\frac{dV}{d\rho}\,u~,
\ee
which is the standard nonlinear Schr\"odinger equation -- recall that one
usually considers $V(\rho)$ as quadratic or cubic in $\rho$. For travelling
waves in the form $u(x,t)=\rho(x-ct)\exp[i\theta(x,t)]$, one finds
solutions to the above nonlinear derivative Schr\"odinger equation that
present velocity restricted to just one sense, and the system is then chiral.
This is nicely illustrated in a more recent work on the same subject
\cite{gse97}, where the soliton structure for vanishing and non-vanishing
boundary conditions are investigated. See also Ref.~{\cite{his97}} for the
case of coupled equations. Evidently, the chiral solitons found in
these works may play important role within the context of the fractional
quantum Hall effect, where chiral excitations are known to appear \cite{jac96}.

On the other hand, some very recent works have shown that chiral solitons
also appear in generalized KdV equations \cite{bmo98}
\be
u_t+f_x-\delta\,u_{xxx}=0~,
\ee
and in generalized Burgers \cite{whi74} equations \cite{baz98}
\be
u_t+f_x-\nu\,u_{xx}=0~.
\ee
Here $f=f(u)$ is some smooth function of $u$ and $f_x=(df/du)\,u_x$ is the
term that introduces nonlinearity. The KdV equation combines dispersion, 
controlled by the real parameter $\delta$, and nonlinearity, described
by $f(u)$. The Burgers equation combines nonlinearity with diffusion 
related to the real parameter $\nu$. In Ref.{\cite{baz98}} we have also
studied the generalized Burgers-Huxley equation, in which extra 
nonlinear terms are introduced: 
\be
u_t+f_x-\nu\,u_{xx}=B\,\frac{df}{du}~.
\ee
Within this context, it seems interesting to extend the presence of chiral
solitons to the new scenario where the KdV and Burguers-Huxley 
equations are added to play the game. The equation that we
consider is given by
\be
\label{eq:}
u_t+f_x+g_{xx}-\delta\,u_{xxx}= h(u) ~,
\ee
where $f$, $g$, and $h$ are smooth functions in $u$. We name this equation the
generalized KdV-Burgers-Huxley or gKdVBH equation. It contains several
interesting particular cases. For $h=0$, it corresponds to the generalized
KdV-Burgers or gKdVB equation \cite{baz98}. 
For $f(u)=(1/2)\lambda\,u^2$, $g(u)=-\nu\,u$, 
and $h(u) = 0$, we get to the standard KdVB equation. 
The KdV and Burgers equations were first added in
Ref.{\cite{joh70}} to describe propagation of waves in liquid-filled elastic
tubes. For $f(u)=\lambda\,u^3$ and $g$ and $h$ as above, it represents 
the modified KdVB equation \cite{whi74}. For $g$ trivial, i.e. $g=0$, we name
the equation the generalized KdV-Huxley or gKdVH equation since it is similar
to the generalized Burgers-Huxley equation considered in
Ref.\cite{baz98}, but with the diffusion term present in the Burgers-Huxley
case changed by the dispersion term present in the KdV case.

In this work we investigate the above-mentioned nonlinear equations, 
with focus on the search for soliton solutions that present chirality. 
For some appropriate choices of the functions $f(u)$, $g(h)$, and $h(u)$, 
we show that it is possible to map each one of the equations onto 
the equation of motion for localized travelling configurations in
relativistic field theoretical systems of real scalar fields $\phi(x,t)$,
with general Lagrangian density in bidimensional $1+1$ spacetime
given by \cite{bmo98,baz98,raja,baz95,baz96}
\be
{\cal L} = \frac{1}{2} \frac{\partial \phi}{\partial x^{\alpha}} 
\frac{\partial \phi}{\partial x_{\alpha}} - U(\phi) ~,
\ee
where $x^{\alpha} = (x^{0} = t, x^{1} = x)$ and 
$x_{\alpha} = (x^{0} = t, x^{1} = x)$, and $U(\phi)$ is the potential. 
In this context, soliton field configurations propagating with velocity $c$
as $\phi(x,t) = \phi(x - ct) = \phi(y)$ arise \cite{bmo98,baz98} as solutions
of the equation of motion associated with the above Lagrangian density, 
\be
\frac{d^{2}\phi}{dy^{2}} = \frac{\partial U}{\partial \phi} ~.
\ee
In particular, we present suitable choices of functions in Eq.(6) that 
provide mappings of the mentioned nonlinear equations onto 
the $\phi^{4}$ and $\phi^{6}$ field theories, whose solutions present
interesting topological features \cite{bmo98,baz98}. 

The article is organized as follows: in the next section we study the
gKdVH equation. In Sec.~{\ref{sec:gKdVB}} we investigate the gKdVB equation, 
and in Sec.~{\ref{sec:gKdVBH}} we consider the combination of the previous
ones, the gKdVBH equation. In each Section, we analyze both cases of a single
equation and a system of coupled nonlinear equations. 
Finally, conclusions are presented in Sec.~{\ref{sec:comm}}.

\section{The gKdVH Equation}
\label{sec:gKdVH}

\indent
We start by considering the gKdVH equation, i.e., Eq.(6) without the 
term with second order derivative in space 
related to diffusion, $g(u) = 0$, but with nonlinear and dispersion 
effects present: 
\be
u_t+\frac{df}{du}\,u_x-\delta\,u_{xxx}= h(u)~.
\ee
For propagating waves of the type $u(x,t) = u(x - ct) = u(y)$, 
Eq.(9) can be 
solved by the first order equation
\be
\frac{du}{dy}= B h(u) ~,
\ee
if $f(u)=h(u)\,(dh/du)$, $B^{2} = 1/\delta$, 
and the velocity is set by $c = -1/B$, a property that confers chirality  
to the travelling solutions. We now present an explicit example 
in which the solutions are solitons. By considering the case in which
\begin{equation}
h(u) = \lambda (a^{2} - u^{2}) ~,
\end{equation}
which implies, 
\begin{equation}
f(u) = - 2 \lambda ^{2} u( a^{2} - u^{2} ) ~,
\end{equation}
with $\lambda$ and $a$ real parameters, we obtain the 
chiral solitons
\be
u(x,t)= a \tanh[\lambda a B(x-ct-\bar{x})]~,
\ee
where $\bar{x}$ is the center of the solution, which is an arbitrary
point in space. We observe that the gKdVH equation, Eq.(9), 
defined by the choices of 
Eqs.(11) and (12), 
represents the equation of 
motion for localized travelling field configurations of a relativistic 
1 + 1 dimensional $\phi^{4}$ field theory of a real scalar field 
$\phi$ \cite{baz95}. Indeed, one can attest this fact 
by differentiating Eq.(10) with 
respect to $y$, and then making use of Eqs.(11) and (13). The resulting 
equation, when compared with Eq.(8), allows the identification 
of the associated $\phi^{4}$ potential, 
\be
U(\phi)= -\frac{\lambda^{2} B^{2}}{2} \phi^{2} ( 2 a^{2} - \phi^{2} ) ~.
\ee

In this context, studies \cite{baz96} performed on field 
theories of two coupled scalar fields motivate us to generalize 
the above analysis to the case of 
a pair of coupled gKdVH equations,
\ben
u_t+\frac{\partial f}{\partial u}\,u_x+\frac{\partial f}{\partial v}\,v_x-
\delta\,u_{xxx}&=&h(u,v)~,\\
v_t+\frac{\partial\bar{f}}{\partial u}\,u_x+
\frac{\partial\bar{f}}{\partial v}\,v_x-\bar{\delta}\,v_{xxx}&=&\bar{h}(u,v)~.
\een
A similar procedure applied to the travelling waves $u(x,t) = u(y)$ and 
$v(x,t) = v(y)$ 
leads to the pair of first order coupled equations, 
\be
\frac{du}{dy}= B h(u,v) ~,
\ee
\be
\frac{dv}{dy}= \bar{B} \bar{h}(u,v) ~,
\ee
provided that $B^{2} = 1/\delta$, $B = \bar{B}$, $c = -1/B$, and 
\begin{equation}
f(u,v) = h \frac{\partial h}{\partial u} + \bar{h} 
	\frac{\partial h}{\partial v} ~,
\end{equation}
\begin{equation}
\bar{f}(u,v) = \bar{h} \frac{\partial \bar{h}}{\partial v} 
	+ h \frac{\partial \bar{h}}{\partial u} ~.
\end{equation}
As an example also related to the field theory of a pair of coupled 
scalar fields interacting through a potential up to the forth power 
in the fields \cite{baz96}, we consider
\begin{equation}
h(u,v) = \frac{1}{B} ( \lambda - \lambda u^{2} - \mu v^{2} ) ~,
\end{equation}
and
\begin{equation}
\bar{h}(u,v) = - \frac{2 \mu u v}{\bar{B}} ~,
\end{equation}
which lead to,
\begin{equation}
f(u,v) = -2\lambda^{2} c^{2}u(1 - u^{2}) + 2 \mu c^{2}
u v^{2} (\lambda + 2 \mu) ~,
\end{equation}
\begin{equation}
\bar{f}(u,v) = -2\mu \lambda c^{2} v(1 - u^{2}) + 
2 \mu^{2} c^{2} v (v^{2} + 2 u^{2}) ~,
\end{equation}
with coupled chiral solutions for $\lambda / \mu > 2$ given by
\begin{equation}
u(x,t) = \tanh[ 2\mu (x - ct - \bar{x}) ] ~,
\end{equation}
\begin{equation}
v(x,t) = \sqrt{ \frac{\lambda}{\mu} - 2 } \; \sec \mbox{h} 
[ 2 \mu  (x - ct - \bar{x}) ] ~.
\end{equation}

\section{The gKdVB Equation}
\label{sec:gKdVB}

\indent
Let us now focus on the gKdVB equation in which nonlinearity, diffusion, and 
dispersion are present, Eq.(6) with $h(u) = 0$:
\be
u_t+\frac{df}{du}\,u_x+
\frac{d^2g}{du^2}\,u_x^2+\frac{dg}{du}\,u_{xx}-\delta\,u_{xxx}=0~.
\ee
Travelling
solutions $u(y)$ are obtained from
\be
\left(-c+\frac{df}{du}\right)u_y+\frac{d^2g}{du^2}\,u_y^2+
\frac{dg}{du}\, u_{yy}-\delta\,u_{yyy}=0~.
\ee
In the case of $f(u)=-f(-u)$ and $g(u)=-g(-u)$, the gKdVB equation 
is invariant under the discrete symmetry $u\to-u$, so that 
a trivial integration leads to
\be
\frac{dg}{du}\,u_y-\delta\,u_{yy}=c\,u-f(u)~,
\ee
if we impose that the resulting equation mantain the original symmetry. 
Now, for $f(u)=A\,u$, with $c=A$, Eq.(29) is solved by
\be
\frac{du}{dy}=\frac{1}{\delta}\,g(u)~,
\ee
such as found for the gKdVH equation. To introduce a distinct example, let us 
consider now that
\begin{equation}
g(u)=\lambda\,u\,(a^2-u^2) ~,
\end{equation}
which leads to the 
chiral solution
\be
u(x,t)=\sqrt{
(a^2/2)\{ 1+\tanh[(\lambda/\delta) a^2 (x-ct-\bar{x})] \} 
}~,
\ee
since its velocity is just $c=A$. In the context
of field theoretical representations, these choices for $f(u)$ and $g(u)$
correspond to a $\phi^{6}$ theory of real scalar fields \cite{baz95}, 
with potential given by,
\be
U(\phi) = \frac{\lambda^{2}}{2 \delta^{2}} 
\phi^{2} (a^{2} - \phi^{2})^{2} ~.
\ee

Furthermore, as in the previous case we can also extend the analysis to a
system of coupled gKdVB equations in the form,
\ben
u_t+f_x+g_{xx}-\delta\,u_{xxx}&=&0~,\\
v_t+\bar{f}_x+\bar{g}_{xx}-\bar{\delta}\,v_{xxx}&=&0~.
\een
Here $f=f(u,v)$ and $g=g(u,v)$ are odd in $u$ and even in $v$, and 
$\bar{f}=\bar{f}(u,v)$ and $\bar{g}=\bar{g}(u,v)$ are even in $u$ and odd
in $v$, in order to preserve the symmetries in the $(u,v)$ space of the 
original equations. These smooth functions allow us to write the above
equations in the form
\ben
u_t+\frac{\partial f}{\partial u}\,u_x+\frac{\partial f}{\partial v}\,v_x+
\frac{\partial g}{\partial u}\,u_{xx}+\frac{\partial g}{\partial v}\,v_{xx}+
\frac{\partial^2g}{\partial u^2}\,u_x^2+\nonumber\\
2\frac{\partial^2g}{\partial u\partial v}u_x\,v_x+
\frac{\partial^2g}{\partial v^2}\,v_x^2-\delta\,u_{xxx}&=&0~,\\
v_t+\frac{\partial\bar{f}}{\partial u}\,u_x+\frac{\partial\bar{f}}{\partial v}
\,v_x+\frac{\partial\bar{g}}{\partial u}\,u_{xx}+
\frac{\partial\bar{g}}{\partial v}\,v_{xx}+
\frac{\partial^2\bar{g}}{\partial u^2}\,u_x^2+\nonumber\\
2\frac{\partial^2\bar{g}}{\partial u\partial v}u_x\,v_x+
\frac{\partial^2\bar{g}}{\partial v^2}\,v_x^2-\bar{\delta}\,v_{xxx}&=&0~.
\een
For travelling waves $u(y)$ and $v(y)$ we obtain, after integrating them once,
\ben
\frac{\partial g}{\partial u}\,\frac{du}{dy}+
\frac{\partial g}{\partial v}\,\frac{dv}{dy}-\delta\frac{d^2u}{dy^2}&=&
c\,u-f(u,v)~,\\
\frac{\partial\bar{g}}{\partial u}\,\frac{du}{dy}+
\frac{\partial\bar{g}}{\partial v}\,\frac{dv}{dy}-
\bar{\delta}\frac{d^2v}{dy^2}&=&c\,v-\bar{f}(u,v)~.
\een

In order to present an example of coupled chiral soliton solutions 
of Eqs.(38) and (39), we consider for instance,
\begin{equation}
\frac{dg}{dy} = au + bv^{2} ~,
\end{equation}
and
\begin{equation}
\frac{d\bar{g}}{dy} = \bar{a}u^{2} + \bar{b}v ~,
\end{equation}
along with,
\begin{equation}
f(u,v) = -cu + cu^{3} + c \frac{\delta}{\bar{\delta}}  
( 1 + 2 \frac{\delta}{\bar{\delta}} ) u v^{2} - bv^{2}~,
\end{equation}
and
\begin{equation}
\bar{f}(u,v) = -cv + c \frac{\delta}{\bar{\delta}}v^{3} + 
c ( 1 + 2 \frac{\delta}{\bar{\delta}} ) u^{2} v - \bar{a}u^{2} ~.
\end{equation}
By substituting Eqs.(40)-(43) in Eqs.(38) and (39), 
we obtain after setting 
$c = a = \bar{b}$, the following 
system of coupled differential equations:
\begin{equation}
\frac{\delta}{c}\frac{d^{2}u}{dy^{2}} = - ( 1 - u^{2}) u 
+ \frac{\delta}{\bar{\delta}}( 1 + 2 \frac{\delta}{\bar{\delta}} ) 
u v^{2} ~, 
\end{equation}
\begin{equation}
\frac{\bar{\delta}}{c}\frac{d^{2}v}{dy^{2}} = - ( 1 -  
\frac{\delta}{\bar{\delta}} v^{2} ) v + 
( 1 + 2 \frac{\delta}{\bar{\delta}} ) u^{2} v ~, 
\end{equation}
which can also be seen as the equations of motion for localized travelling 
configurations of a relativistic field theory of two coupled scalar fields 
with fourth order potential. 
Eqs.(44) and (45) present solutions given by
\begin{equation}
u(x,t) = \tanh [ 2 (\frac{c\delta}{2 \bar{\delta}^{2}})^{1/2} 
	(x - c t - \bar{x}) ] ~,
\end{equation}
\begin{equation}
v(x,t) = ( \frac{\bar{\delta}}{\delta} - 2 )^{1/2} \sec\mbox{h} 
[ 2 (\frac{c\delta}{2 \bar{\delta}^{2}})^{1/2}
	(x - c t - \bar{x}) ] ~,
\end{equation}
with $\bar{\delta} > 2\delta$, and chiral behavior related to the 
identification of the velocity $c$ with the parameters of the functions 
$f$ and $g$.

\section{The gKdVBH Equation}
\label{sec:gKdVBH}

\indent
At last, we generalize the previous results to the gKdVBH equation, Eq.(6), 
in which extra nonlinear terms are included with respect to the 
gKdVB equation, Eq.(27), through the function $h(u)$. 
By considering travelling solutions of Eq.(6), we obtain
\be
\left(-c+\frac{d\bar{f}}{du}\right)u_y+\frac{d^2g}{du^2}\,u_y^2+
\frac{dg}{du}\, u_{yy}-\delta\,u_{yyy}= h(u)~.
\ee
At this point, by assuming in Eq.(48) the presence of 
symmetry under $u\to-u$, then the function $h(u)$ 
must also have odd parity, as discussed in Sec.3 for the functions $f(u)$ and 
$g(u)$. After a trivial integration, 
the gKdVBH equation can be mapped onto the 
gKdVB equation, Eq.(28), if 
\be 
f(u) = \bar{f}(u) - \int^{y} h[u(y')] dy' ~.
\ee
Therefore, from the analysis presented in Sec.3 we conclude that 
chiral solitons can also be found as solutions of 
the gKdVBH equation, as well as of a pair of coupled gKdVBH equations.
In addition, mappings of Eq.(6) onto field theories of 
real scalar coupled fields are also possible.

\section{Conclusion}
\label{sec:comm}

\indent
In conclusion, we have investigated and introduced explicit examples of
chiral soliton solutions in generalized equations such as the gKdVH,
the gKdVB and the gKdVBH equations. These nonlinear partial differential
equations combine diffusion, dispersion, and nonlinearity in distinct ways.
We have also shown that they can be maped to  equations of motion that 
appear in relativistic systems of real scalar fields, provided 
appropriate choices of the functions that define them are made. 
In particular, we illustrate this point by presenting examples related to
$\phi^{4}$ and $\phi^{6}$ field theories. In each case considered, 
the velocity of the solutions is determined in terms of the parameters
of the nonlinear equation, therefore characterizing their chiral feature. 
We have also found chirality in systems of two coupled equations, which 
were also shown to be mapped onto field theories of pairs 
of real scalar coupled fields.

\section*{Acknowledgments} 

DB and EPR would respectively 
like to thank the Center for Theoretical Physics at 
Massachusetts Institute of Technology and the Condensed
Matter Theory group at Harvard University 
for hospitality.

\end{document}